\documentclass{article}
\usepackage{spconf,amsmath,graphicx,hyperref}
\usepackage{subcaption}
\usepackage{amssymb} 
\usepackage[subtle,title=normal,sections=normal,margins=normal,mathdisplays=normal,mathspacing=normal, lists=normal]{savetrees}
\usepackage{booktabs} 
\usepackage{amsmath}  
\usepackage{graphicx} 
\usepackage{balance}
\usepackage{cite}

\ninept

\DeclareMathOperator*{\argmax}{argmax}
\DeclareMathOperator*{\argmin}{argmin}

\title{Reference Microphone Selection for Guided Source Separation \\ based on the Normalized L-p Norm}
\name{Anselm Lohmann$^1$, Tomohiro Nakatani$^2$, Rintaro Ikeshita$^2$, Marc Delcroix$^2$, Shoko Araki$^2$, Simon Doclo$^1$}
\address{$^{1}$Carl von Ossietzky Universität Oldenburg, Dept. of Medical Physics and Acoustics, Germany \\
$^{2}$NTT, Inc., Japan\\
{\tt anselm.lohmann@uni-oldenburg.de}
}
\begin{document}
%
\maketitle
\begin{abstract}
Guided Source Separation (GSS) is a popular front-end for distant automatic speech recognition (ASR) systems using spatially distributed microphones. When considering spatially distributed microphones, the choice of reference microphone may have a large influence on the quality of the output signal and the downstream ASR performance. In GSS-based speech enhancement, reference microphone selection is typically performed using the signal-to-noise ratio (SNR), which is optimal for noise reduction but may neglect differences in early-to-late-reverberant ratio (ELR) across microphones.  In this paper, we propose two reference microphone selection methods for GSS-based speech enhancement that are based on the normalized $\ell_p$-norm, either using only the normalized $\ell_p$-norm or combining the normalized $\ell_p$-norm and the SNR to account for both differences in SNR and ELR across microphones. Experimental evaluation using a CHiME-8 distant ASR system shows that the proposed $\ell_p$-norm-based methods outperform the baseline method, reducing the macro-average word error rate.
\end{abstract}
\section{Introduction}
\label{sec:intro}
\begin{figure}[t!]
\centering
\includegraphics[width=0.95\columnwidth]{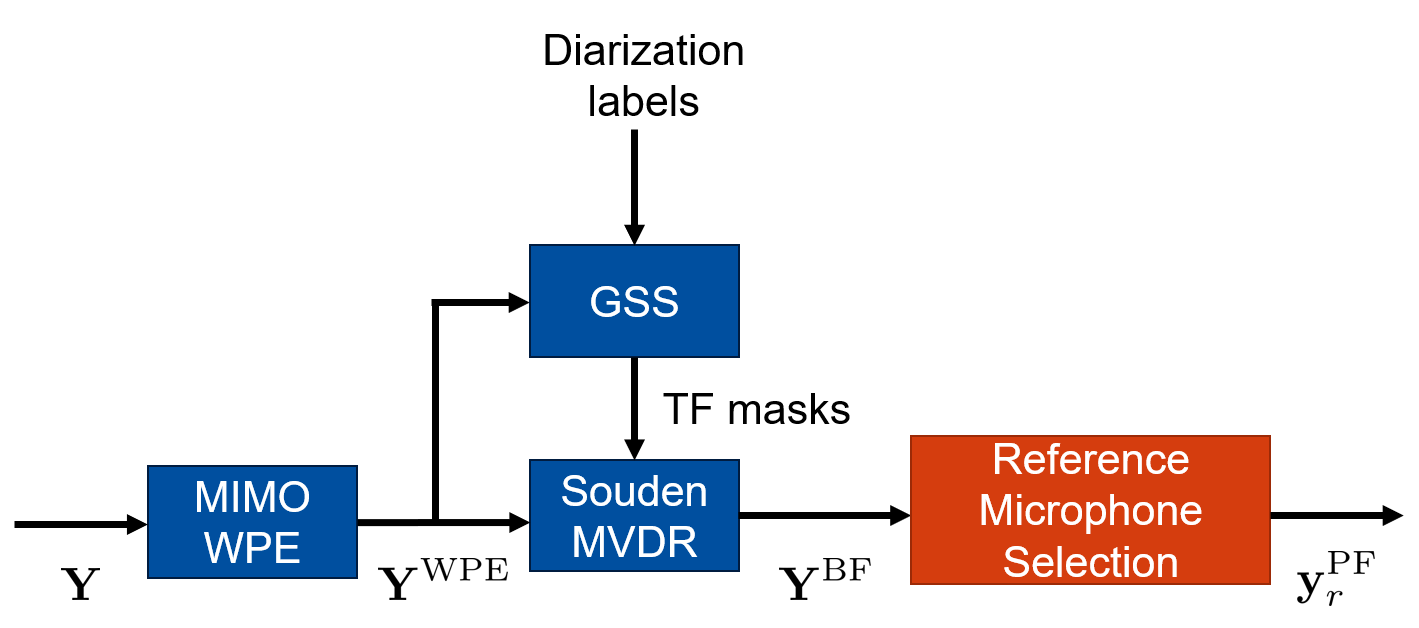}
\vspace*{-0.40cm}
\caption{GSS-based speech enhancement}
\label{fig:gss_diagram}
\vspace{-0.55cm}
\end{figure}

Guided source separation (GSS)-based speech enhancement is a popular approach \cite{Boeddecker2018} for enhancing a target speech source in noisy and reverberant environments, particularly in the context of the CHiME challenge, which aims at ASR and diarization of multi-talker conversations recorded by spatially distributed microphones \cite{Cornell2024,Park2023,Mitrofanov2024,Kamo2026}. GSS-based speech enhancement (see Fig. \ref{fig:gss_diagram}) first performs dereverberation using the multiple-input multiple-output (MIMO) weighted prediction error (WPE) dereverberation method \cite{Yoshioka2012}. In a second step, noise reduction is performed using a MIMO minimum variance distortionless response (MVDR) beamformer \cite{Souden2010}, where the target speech and noise covariance matrices are estimated with time-frequency masks computed using GSS \cite{Boeddecker2018}. In order to output an enhanced target source signal, the GSS-based approach selects the reference microphone index of the MIMO beamformer with the highest estimated output signal-to-noise ratio (SNR) \cite{LawinOre2012}.

When performing speech enhancement using spatially distributed microphones, e.g. GSS-based, there may be large differences in the early-to-late reverberation ratio (ELR) and SNR in each microphone. Hence, the choice of the reference microphone may have a large influence on the quality of the output signal and downstream ASR performance \cite{Cornell2021,LawinOre2012,Araki2018,Zhang2021, Lohmann2024}. While different reference microphone selection methods have been proposed for noise reduction, e.g. selecting the microphone with the largest input signal power of the target speech source \cite{LawinOre2012, Araki2018}, selecting the microphone with the highest output SNR, as in GSS-based speech enhancement, is considered optimal \cite{LawinOre2012, Zhang2021}. However, for WPE dereverberation, it was recently proposed to select the microphone with the lowest output normalized $\ell_p$-norm \cite{Lohmann2024}. Hence, since GSS-based speech enhancement includes both WPE dereverberation and noise reduction, selecting the reference microphone optimal for noise reduction may not result in selecting the microphone with the highest overall signal quality and ASR performance.

In this paper, we propose reference microphone selection methods based on the normalized $\ell_p$-norm for GSS-based speech enhancement. The normalized $\ell_p$-norm \cite{Hurley2009} measures the sparsity of a signal in the time-frequency domain and was shown to typically select a microphone with a high input ELR in noise-free conditions \cite{Lohmann2024}. In order to account for differences in input ELR between the microphones, the first proposed reference microphone selection method uses only the normalized $\ell_p$-norm of the MIMO beamformer output signals. However, since this does not take into account differences in output SNR between microphones, the second proposed reference microphone selection method combines the normalized $\ell_p$-norm and SNR of the beamformer output. Experimental evaluation using signal quality metrics show that using only the normalized $\ell_p$-norm significantly outperforms using only the SNR at high input SNRs, while using both the normalized $\ell_p$-norm and SNR consistently outperforms using only the SNR. Experimental evaluation using a CHiME-8 distant ASR system \cite{Cornell2024} shows that the proposed $\ell_p$-norm-based reference microphone selection methods outperform the baseline method using only the SNR in terms of word error rate (WER), with the combination of the normalized $\ell_p$-norm and SNR yielding the lowest WER.
\vspace*{-0.2cm}
\section{GSS-based Speech Enhancement}
We consider a scenario where $K$ speech sources are recorded in a reverberant and noisy environment by $M$ spatially distributed microphones, with $K<M$. In the short-time Fourier transform (STFT) domain, let $f\in\{1,...,F\}$ be the frequency-bin index and $l\in\{1,...,L\}$ be the time-frame index. The reverberant and noisy mixture vector in the $m$-th microphone $\mathbf{y}_m(f) = \begin{bmatrix} y_m(f,1) & \cdots & y_m(f,L)\end{bmatrix}^T\in\mathbb{C}^{L}$, with $(.)^T$ denoting the transpose operator, can be written as 
\begin{equation}
\mathbf{y}_m(f) = \mathbf{x}^{\text{early}}_m(f) + \mathbf{x}^{\text{late}}_m(f)+\mathbf{n}_m(f),
\label{eq:vec_sigmodel}
\end{equation}
where $\mathbf{x}^{\text{early}}_m\in\mathbb{C}^{L}$ and $\mathbf{x}^{\text{late}}_m\in\mathbb{C}^{L}$ denote the early-reverberant and late-reverberant target speech signal vectors, respectively, and $\mathbf{n}_m\in\mathbb{C}^{L}$ denotes the noise mixture vector, consisting of background noise and $K-1$ reverberant interfering speech signals. The equation in \eqref{eq:vec_sigmodel} can be rewritten by stacking the signal and mixture vectors across microphones, i.e.
\begin{equation}
\mathbf{Y}(f) = \mathbf{X}^{\text{early}}(f) + \mathbf{X}^{\text{late}}(f)+\mathbf{N}(f),
\label{eq:vec_sigmodel}
\end{equation}
where $\mathbf{Y}(f) = \begin{bmatrix} \mathbf{y}_1(f) & \cdots &\mathbf{y}_M(f) \end{bmatrix}^T\in\mathbb{C}^{M\times L}$ denotes the reverberant and noisy mixture matrix, $\mathbf{X}^{\text{early}}(f)\in\mathbb{C}^{M\times L}$ and $\mathbf{X}^{\text{late}}(f)\in\mathbb{C}^{M\times L}$ denote the early-reverberant and late-reverberant target speech signal matrices, respectively, and $\mathbf{N}(f)\in\mathbb{C}^{M\times L}$ denotes the noise mixture matrix. For clarity, the frequency-bin index $f$ will be omitted where possible.

First, GSS-based speech enhancement in Fig. \ref{fig:gss_diagram} performs dereverberation using MIMO WPE, i.e. by subtracting an estimate of the late-reverberant target signal, alongside the late-reverberant intereferer signals, from the reverberant and noisy mixture. The MIMO WPE output mixture matrix $\mathbf{Y}^{\text{WPE}}$ can be written as
\begin{equation}
\mathbf{Y}^{\text{WPE}} = \mathbf{Y} - \mathbf{G}^H\tilde{\mathbf{Y}}_{\tau},
\end{equation}
where $\mathbf{G}\in\mathbb{C}^{ML_g\times M}$ denotes the MIMO WPE filter with filter length $L_g$, $(.)^H$ denotes the Hermitian transpose operator and $\tilde{\mathbf{Y}}_{\tau}\in\mathbb{C}^{ML_g\times L}$ is a multi-channel convolution matrix of the delayed reverberant and noisy mixture with $\tau$ the prediction delay \cite{Yoshioka2012}. The MIMO WPE filter can be computed by minimizing \cite{Jukic2015mimo} 
\begin{equation}
J_{\ell_{\mathbf{\Phi};p,2}}(\mathbf{Y}^{\text{WPE}}) = \| \mathbf{Y}^{\text{WPE}} \|_{\mathbf{\Phi};2,p}^p = \sum_{l=1}^L\lvert\left(\mathbf{y}^{\text{WPE}}_{1:M}(l)\right)^H\mathbf{\Phi}^{-1}\mathbf{y}^{\text{WPE}}_{1:M}(l)\rvert^{\frac{p}{2}},
\label{eq:mimo_wpe_cost}
\end{equation}
where $J_{\ell_{\mathbf{\Phi};p,2}}(\mathbf{Y}^{\text{WPE}})$ denotes the mixed norm $\ell_{\mathbf{\Phi};p,2}$ of the WPE output mixture matrix $\mathbf{Y}^{\text{WPE}}$ with group matrix $\mathbf{\Phi}\in\mathbb{C}^{M\times M}$ and sparsity-promoting parameter $p$ \cite{Jukic2015mimo} and $\mathbf{y}^{\text{WPE}}_{1:M}(l)\in\mathbb{C}^{M}$ denotes the $l$-th column vector of $\mathbf{Y}^{\text{WPE}}$. Typically, the cost function in \eqref{eq:mimo_wpe_cost} is minimized using the iteratively reweighted least squares algorithm for $I_{\text{WPE}}$ iterations \cite{Yoshioka2012}\cite{Jukic2015mimo}.

In a second step, GSS-based speech enhancement performs noise reduction using the Souden MVDR beamformer \cite{Souden2010} on the MIMO WPE output mixture matrix $\mathbf{Y}^{\text{WPE}}$. The MIMO beamformer output signal matrix $\mathbf{Y}^{\text{BF}}$ can be written as
\begin{equation}
\mathbf{Y}^{\text{BF}} = \mathbf{W}^H\mathbf{Y}^{\text{WPE}},
\end{equation}
with the MIMO beamformer filter $\mathbf{W} = \begin{bmatrix} \mathbf{w}_1 & \cdots & \mathbf{w}_M \end{bmatrix}\in\mathbb{C}^{M\times M}$ computed as
\begin{equation}
\mathbf{W} = \frac{\hat{\mathbf{R}}_{n}^{-1}\hat{\mathbf{R}}_{x}}{\text{Tr}(\hat{\mathbf{R}}_{n}^{-1}\hat{\mathbf{R}}_{x})}
\end{equation}
using estimated batch covariance matrices $\hat{\mathbf{R}}_{n}$ and $\hat{\mathbf{R}}_{x}$ of the noise mixture and early-reverberant target speech signal matrices, respectively.  
The covariance matrices $\hat{\mathbf{R}}_{n}$ and $\hat{\mathbf{R}}_{x}$ are computed using time-frequency masks for the early-reverberant target source signal and noise mixture, i.e.
\begin{equation}
\hat{\mathbf{R}}_{n} = \frac{1}{L}\sum_{l=1}^L\mu_n(l)\mathbf{y}^{\text{WPE}}_{1:M}(l)\left(\mathbf{y}^{\text{WPE}}_{1:M}(l)\right)^H,
\end{equation}

\begin{equation}
\hat{\mathbf{R}}_{x} = \frac{1}{L}\sum_{l=1}^L\mu_{x}(l)\mathbf{y}^{\text{WPE}}_{1:M}(l)\left(\mathbf{y}^{\text{WPE}}_{1:M}(l)\right)^H,
\end{equation}
where $\mu_x(l)\in\mathbb{R}$ denotes the (early-reverberant) target signal time-freqency mask and $\mu_n(l)\in\mathbb{R}$ denotes the noise mixture time-frequency mask, such that $\mu_x(l) + \mu_n(l) = 1$. The time-frequency masks $\mu_x(l)$ and $\mu_n(l)$ are computed using GSS \cite{Boeddecker2018} by assuming a complex angular central Gaussian mixture model (cACGMM) \cite{Ito2016} for the $K$ speech signals and background noise signal, i.e
\begin{equation}
\mathcal{P}\left(\bar{\mathbf{y}}_{1:M}^{\text{WPE}}(l); \phi_k,\mathbf{B}_k\right) = \sum_{k=1}^{K+1} \frac{\pi^{-M}\phi_k(M-1)!}{2\text{det}(\mathbf{B}_k)\left((\bar{\mathbf{y}}_{1:M}^{\text{WPE}}(l))^H\mathbf{B}^{-1}_k\bar{\mathbf{y}}_{1:M}^{\text{WPE}}(l)\right)^M} ,
\label{eq:cACGMM_probability}
\end{equation}
where $\mathcal{P}\left(\bar{\mathbf{y}}_{1:M}^{\text{WPE}}(l); \phi_k,\mathbf{B}_k\right)$ denotes the probability distribution of the normalized WPE output mixture $\bar{\mathbf{y}}_{1:M}^{\text{WPE}}(l) = \frac{\mathbf{y}_{1:M}^{\text{WPE}}(l)}{\lVert \mathbf{y}_{1:M}^{\text{WPE}}(l) \rVert_2}$ given the prior probability of the $k$-th, $k\in\{1,...,K+1\}$, speech and background noise signal $\phi_k$ and the cACGMM concentration matrix of the $k$-th signal $\mathbf{B}_k\in\mathbb{C}^{M\times M}$. By maximizing the log-likelihood function of \eqref{eq:cACGMM_probability}, typically using the expectation-maximization algorithm for $I_{\text{GSS}}$ iterations \cite{Boeddecker2018}, the time-frequency masks $\mu_x(l)$ and $\mu_n(l)$ can be computed using the posterior probability of the $K+1$ signals. When computing the posterior probability, GSS effectively uses source activity information provided by diarization labels \cite{Boeddecker2018}.

Finally, a reference microphone index of the MIMO beamformer $r\in\{1,...,M\}$ is selected, such that the output is a single channel $\mathbf{y}_r^{\text{BF}}$, after which the blind analytic postfilter is applied \cite{Boeddecker2018}\cite{Warsitz2007}, i.e. 
\begin{equation}
\mathbf{y}_r^{\text{PF}} = \frac{\lvert\mathbf{w}^H_r\hat{\mathbf{R}}^2_n\mathbf{w}_r\rvert^{\frac{1}{2}}}{\mathbf{w}_r^H\hat{\mathbf{R}}_n\mathbf{w}_r}\mathbf{y}_r^{\text{BF}}.
\end{equation}
The baseline reference microphone selection method as well as the proposed normalized $\ell_p$-norm-based reference microphone methods will be discussed in Section 3.
\vspace{-0.2cm}
\section{Reference Microphone Selection}
\label{sec:mic_selection}
Section 3.1 describes the baseline reference microphone selection method for GSS-based speech enhancement, selecting the beamformer output with the highest estimated SNR \cite{Boeddecker2018}. While it may be optimal in terms of noise reduction, it may not result in the highest overall signal quality or ASR performance, as it does not take into account differences in ELR between the microphone signals. Therefore, in Sections \ref{sec:nlp_refmicsel} and \ref{sec:nlp_snr_refmicsel}, we propose reference microphone selection methods based on the normalized $\ell_p$-norm of the beamformer output.
\vspace{-0.2cm}
\subsection{Baseline microphone selection using SNR}
\label{sec:snr_micsel}
In \cite{LawinOre2012, Zhang2021}, it has been proposed to perform reference microphone selection by selecting the MIMO beamformer output with the highest estimated broadband SNR, i.e.
\begin{equation}
r_{\text{SNR}} = \argmax_{m}J_{\hat{\text{SNR}}}(\mathbf{y}_m^{\text{BF}}),
\label{eq:snr_refmicsel}
\end{equation}
where $J_{\hat{\text{SNR}}}(\mathbf{y}_m^{\text{BF}})$ denotes the estimated SNR in the beamformer output. Alternatively, the reference microphone selection problem in \eqref{eq:snr_refmicsel} can also be formulated as a minimizing the noise-to-signal ratio
where $J_{\hat{\text{NSR}}}(\mathbf{y}_r^{\text{BF}})=1/J_{\hat{\text{SNR}}}(\mathbf{y}_r^{\text{BF}})$.
Given the target source and noise covariances matrices $\hat{\mathbf{R}}_{x}$ and $\hat{\mathbf{R}}_{n}$ and the filter $\mathbf{w}_m$, the estimated broadband SNR of the $m$-th beamformer output can be computed as
\begin{equation}
J_{\hat{\text{SNR}}}(\mathbf{y}_m^{\text{BF}}) = \frac{\sum_{f=1}^{F} \mathbf{w}_m^H(f) \hat{\mathbf{R}}_{x}(f) \mathbf{w}_m(f)}{\sum_{f=1}^{F} \mathbf{w}_m^H(f)\hat{\mathbf{R}}_{n}(f) \mathbf{w}_m(f)}.
\end{equation}
\subsection{Microphone selection using normalized $\ell_p$-norm}
\label{sec:nlp_refmicsel}
In \cite{Lohmann2024}, the reference microphone selection problem for multiple-input single-output WPE dereverberation was formulated using the $\ell_p$-norm cost function of the WPE output signals. However, since the $\ell_p$-norm depends on signal power, it was proposed to use the $\ell_p$-norm of the power-normalized WPE output signals, known as the normalized $\ell_p$-norm \cite{Lohmann2024}\cite{Hurley2009}, i.e. $J_{\ell_p/\ell_2}(\mathbf{z}) = \sum_{f=1}^{F}\frac{\left\| \mathbf{z}(f)\right\|_{p}}{\left\|\mathbf{z}(f)\right\|_{2}}$. Instead of applying the normalized $\ell_p$-norm directly to the WPE output, in order to mitigate the effect of noise, we propose to perform reference microphone selection for GSS-based speech enhancement by selecting the beamformer output with the lowest normalized $\ell_p$-norm, i.e.
\begin{equation}
r_{\ell_p/\ell_2} = \argmin_{m}J_{\ell_p/\ell_2}(\mathbf{y}_m^{\text{BF}}) = \argmin_{m}\sum_{f=1}^{F}\frac{\left\| \mathbf{y}^{\text{BF}}_{m}(f)\right\|_{p_{\epsilon}}}{\left\|\mathbf{y}^{\text{BF}}_{m}(f)\right\|_{2}},
\label{eq:nlp_refmicsel}
\end{equation}
where $p_{\epsilon} = \text{max}\{p,\epsilon\}$ is used to avoid numerical issues for $p=0$ with $\epsilon$ a small constant.
Using \eqref{eq:nlp_refmicsel}, the beamformer output with the sparsest time-frequency representation is selected, typically corresponding to a microphone with a high input ELR \cite{Lohmann2024}. However, since noise and interfering sources also degrade sparsity in the time-frequency domain \cite{Bofill2000}, this reference microphone selection method may also inherently favor a beamformer output where the target source is most prominent.
\vspace{-0.1cm}
\subsection{Microphone selection using SNR and normalized $\ell_p$-norm}
\label{sec:nlp_snr_refmicsel}
In order to perform reference microphone selection using both the SNR and the normalized $\ell_p$-norm of the beamformer output $\mathbf{y}_r^{\text{BF}}$, we straightforwardly combine the estimated noise-to-signal ratio with the normalized $\ell_p$-norm after scaling them to the same range, i.e.
\begin{equation}
r_{\text{comb}} = \argmin_m \alpha\tilde{J}_{\ell_p/\ell_2}(\mathbf{y}_m^{\text{BF}}) + (1 - \alpha)\tilde{J}_{\hat{\text{NSR}}}(\mathbf{y}_m^{\text{BF}}),
\label{eq:combined_problem}
\end{equation}
where $\alpha \in [0, 1]$ is a trade-off parameter and $\tilde{J}_{\hat{\text{NSR}}}(\mathbf{y}_m^{\text{BF}})$ and $\tilde{J}_{\ell_p/\ell_2}(\mathbf{y}_m^{\text{BF}})$ denote the scaled noise-to-signal ratio and normalized $\ell_p$-norm after min-max normalization, i.e.
\begin{equation}
\tilde{J}(\mathbf{y}_m^{\text{BF}}) = \frac{J(\mathbf{y}_m^{\text{BF}}) - \min_m J(\mathbf{y}_m^{\text{BF}})}{\max_m J(\mathbf{y}_m^{\text{BF}}) - \min_m J(\mathbf{y}_m^{\text{BF}})}.
\end{equation}
  By applying min-max normalization, the values are normalized across microphones, assigning a value of 0 to the microphone with the lowest original value and a value of 1 to the microphone with the highest original value. By scaling the values of the normalized $\ell_p$-norm and noise-to-signal ratio in this way, the differences in their original ranges are removed, ensuring that the trade-off parameter $\alpha$ can effectively balance their respective contributions. Hence, the proposed method trades off between using only the SNR and using only the normalized $\ell_p$-norm, such that a microphone with both a high input ELR and high output SNR may be selected, which may lead to a higher overall signal quality and ASR performance. Note that for $\alpha=0$, the proposed method corresponds to using only the SNR in \eqref{eq:snr_refmicsel}, while for $\alpha = 1$ the method corresponds to using only the normalized $\ell_p$-norm in \eqref{eq:nlp_refmicsel}.

\section{Experimental evaluation}
\label{sec:eval}
In this section, we evaluate the performance of the proposed reference microphone selection methods for GSS-based speech enhancement. In Section \ref{sec:gss_implementation}, we briefly describe the parameters used to evaluate the proposed methods. In Section \ref{sec:simulation_results}, we evaluate the signal quality in the selected reference microphones for the baseline and proposed methods using non-intrusive signal quality metrics on simulated data. After confirming the signal quality improvements over the baseline method, in Section \ref{sec:chime8_asr}, we evaluate the ASR performance of the proposed reference microphone selection methods by integrating them into a CHiME-8 distant ASR system. First, we set the value of the trade-off parameter used in all further experiments for the combined method on the CHiME-8 development data. Then, we compare the performance of the baseline and the proposed methods on the the CHiME-8 evaluation data. 
\vspace{-0.1cm}
\subsection{Implementation parameters}
\label{sec:gss_implementation}
For all experiments, GSS-based speech enhancement was performed with an STFT framework, MIMO WPE, Souden MVDR beamformer and GSS parameters identical to \cite{Park2023}. The STFT framework used a sampling rate of $16000$ Hz with an STFT frame length of $64$ ms, a frame shift of $16$ ms and a Hanning window. MIMO WPE was implemented with the sparsity-promoting parameter $p=0$, the group matrix $\mathbf{\Phi} = \mathbf{I}$, the identity matrix, the filter length $L_g = 5$, the prediction delay $\tau = 2$ and $I_{\text{WPE}} = 3$ iterations. As in \cite{Park2023}, post-masking is applied to the beamformer output after reference microphone selection. GSS was implemented using $I_{\text{GSS}} = 5$ iterations. The small constant for the normalized $\ell_p$-norm was set to $\epsilon=10^{-4}$.
\vspace{-0.1cm}
\subsection{Signal quality on simulated data}
\label{sec:simulation_results}
We considered 3 arrays of 4 closely-spaced microphones distributed in a reverberant room of dimensions $7 \text{ m} \times 7 \text{ m} \times 2.5 \text{ m}$ with a randomly chosen reverberation time $T_{60}$ between 200 and 500 ms. The positions of a target source and 3 arrays of 4 closely-spaced microphones were also randomized within this room. A total of 100 unique reverberant impulse responses were simulated using Pyroomacoustics and convolved with clean speech from Librispeech to create reverberant utterances of the target source. For all utterances, background noise was generated at a specific SNR using noise from the CHiME-6 dataset \cite{Cornell2024}. 

We used oracle diarization labels for GSS-based speech enhancement and set the trade-off parameter $\alpha=0.5$ for the combined reference microphone selection method in Section \ref{sec:nlp_snr_refmicsel}.

Due to the inherently large and diverse time-differences of arrival and differences in signal power when using spatially distributed microphones, we considered non-intrusive signal metrics \cite{Shi2025}: DNSMOS \cite{Reddy2022}, NISQA (MOS) \cite{Mittag2021}, SCOREQ \cite{Ragano2024}, non-intrusive PESQ (NI-PESQ) \cite{Kumar2023} and non-intrusive STOI (NI-STOI) \cite{Kumar2023}. In addition, signal statistics in terms of the estimated SNR of the beamformer output ($\hat{\text{oSNR}}$) in decibel (dB) and ELR in the noisy and reverberant microphone signals (iELR) in dB, defined using a cut-off of 30 ms between early and late reverberation, of the selected reference microphones are included. All reported values have been averaged across all 100 utterances.

Table \ref{tab:snr10} and Table \ref{tab:snr-10} show the signal quality as measured by a wide range of non-intrusive metrics and select signal statistics using the baseline and the proposed reference microphone selection methods given an input SNR of 10 dB and -10 dB, respectively. We will first discuss the overall performance based on these metrics and then analyze the differences in signal statistics.

In terms of signal quality, at 10 dB input SNR in Table \ref{tab:snr10}, using only the normalized $\ell_p$-norm for reference microphone selection achieves a significant improvement compared to using only the SNR for reference microphone selection, whereas using both the SNR and normalized $\ell_p$-norm achieves a smaller improvement. At -10 dB input SNR in Table \ref{tab:snr-10}, using both the normalized $\ell_p$-norm achieves an improvement compared to using only the normalized $\ell_p$-norm or only the SNR. 

In terms of signal statistics, at an input SNR of 10 dB in Table \ref{tab:snr10}, using both the SNR and the normalized $\ell_p$-norm clearly trades-off between the estimated output SNR in the beamformer output and the input ELR in the reference microphone. Meanwhile at an input SNR of -10 dB in Table \ref{tab:snr-10}, while the trend in estimated output SNR follows that in Table \ref{tab:snr10}, the trend in ELR is not as clear, as the performance of the normalized $\ell_p$-norm is degraded in the presence of noise.

\begin{table}[h!]
\centering
\caption{Signal quality measured using non-intrusive metrics and signal statistics of selected reference microphone signals for (a) 10 dB and (b) -10 dB input SNR.}
\label{tab:combined_signal_quality}
\vspace{-0.5cm}
\begin{subtable}{\columnwidth}
    \centering
    \caption{Input SNR of 10 dB}
    \label{tab:snr10}
    \resizebox{\linewidth}{!}{%
    \setlength{\tabcolsep}{2pt} 
    \begin{tabular}{lccccc|ccc}
    \toprule
    Method & DNSMOS & NISQA & SCOREQ & NI-PESQ & NI-STOI & $\hat{\text{oSNR}}$& iELR \\
    \midrule
    SNR & 2.88 & 3.61 & 2.67 & 2.01 & 0.92 & \textbf{24.14} & 7.79 \\
    Normalized $\ell_p$-norm & \textbf{2.94} & \textbf{3.69} & \textbf{2.84} & \textbf{2.25} & \textbf{0.94} & 23.74 & \textbf{9.56} \\
    Combination ($\alpha=0.5$) & 2.92 & 3.67 & 2.76 & 2.14 & 0.93 & 24.04 & 8.76 \\
    \bottomrule
    \end{tabular}%
    }
\end{subtable}

\vspace{0.2cm} 

\begin{subtable}{\columnwidth}
    \centering
    \caption{Input SNR of -10 dB}
    \label{tab:snr-10}
    \resizebox{\linewidth}{!}{%
    \setlength{\tabcolsep}{2pt}
    \begin{tabular}{lccccc|ccc}
    \toprule
    Method & DNSMOS & NISQA & SCOREQ & NI-PESQ & NI-STOI & $\hat{\text{oSNR}}$ & iELR \\
    \midrule
    SNR & 2.09 & 1.97 & 1.75 & \textbf{1.25} & 0.86 & \textbf{19.74} & 8.90 \\
    Normalized $\ell_p$-norm & 2.09 & 1.98 & 1.75 & \textbf{1.25} & 0.86 & 18.82 & 8.90 \\
    Combination ($\alpha=0.5$) & \textbf{2.11} & \textbf{1.99} & \textbf{1.76} & \textbf{1.25} & \textbf{0.87} & 19.39 &  \textbf{9.13}\\
    \bottomrule
    \end{tabular}%
    }
\end{subtable}
\vspace*{-0.35cm}
\end{table}

\subsection{CHiME-8 ASR performance}
\label{sec:chime8_asr}
We evaluate the downstream ASR performance of the proposed reference microphone selection methods by using GSS-based speech enhancement as a front-end for a CHiME-8 distant ASR system \cite{Cornell2024}. In the CHiME-8 distant ASR system, GSS-based speech enhancement is applied after reducing the number of utilized microphones, followed by transcription using ASR \cite{Cornell2024}\cite{Park2023}\cite{Kamo2026}. In addition, a diarization system is also included to estimate the diarization labels for GSS-based speech enhancement \cite{Cornell2024}\cite{Park2023}\cite{Kamo2026}. The ASR system used is a 0.6B
parameter Conformer-based transducer model \cite{Park2023}. The diarization system used is an end-to-end neural diarization with a vector
clustering and multi-channel source counting-based model \cite{Tawara2025}. We considered both oracle diarization labels as well as estimated diarization labels \cite{Tawara2025} for GSS. The ASR performance is measured using the time-constrained minimum-permutation WER (tcpWER).

The CHiME-8 challenge \cite{Cornell2024} requires evaluating the distant ASR system on 4 datasets: CHiME-6 (CH6), DiPCo (DiP), NOTSOFAR1 (NSF) and Mixer 6 (Mi6). The CHiME-6 and DiPCo datasets consist of dinner party scenarios recorded using 6 spatially distributed devices, each containing a linear array of 4 compact microphones and 5 spatially distributed devices, each containing a circular array of 7 microphones, respectively. The Mixer 6 dataset consists of one-on-one interview scenarios recorded using 10 spatially distributed devices, each containing a single microphone. The NOTSOFAR1 dataset used in the CHiME-8 challenge consists of business meeting-like scenarios recorded using a single microphone array, unlike its standalone counterpart \cite{Cornell2024}. Therefore, unlike the other three datasets, it is not recorded using spatially distributed microphones. Note that currently the NOTSOFAR1 dataset is only available in the CHiME-8 evaluation data. The macro-average tcpWER is computed as an average of the tcpWER across all datasets \cite{Cornell2024}.

\begin{figure}[t!]
\centering
\includegraphics[width=0.9\columnwidth]{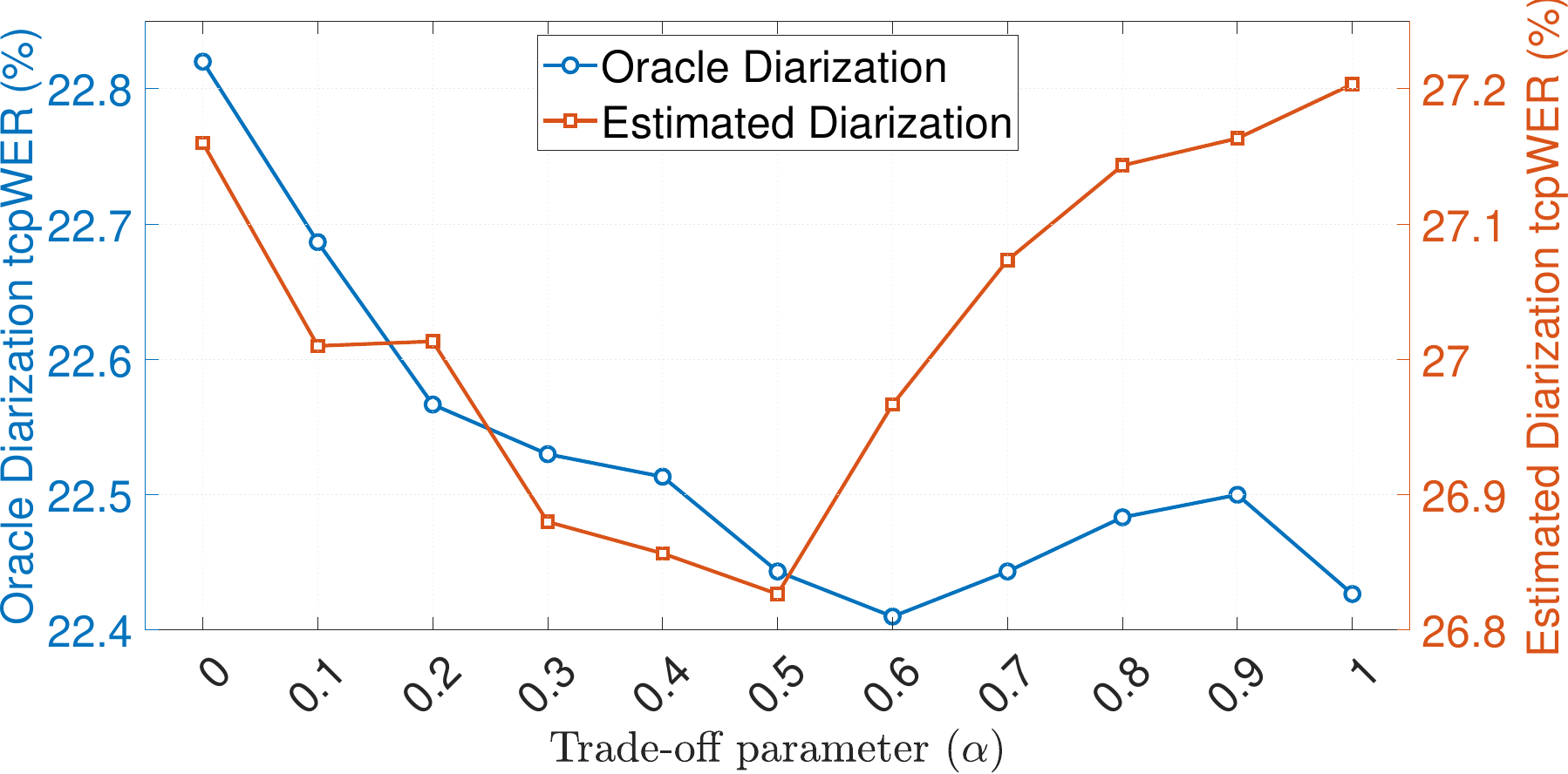}
\caption{ASR performance of CHiME-8 system using both the normalized $\ell_p$-norm and SNR for reference microphone selection in terms of macro-average tcpWER (\%) using oracle diarization labels and estimated diarization labels on CHiME-8 development data for different values of the trade-off parameter $\alpha$}
\label{fig:setalpha_dev}
\vspace{-0.6cm}
\end{figure}

Fig. \ref{fig:setalpha_dev} shows the ASR performance of the CHiME-8 system with the proposed method using both the SNR and normalized $\ell_p$-norm on the CHiME-8 development data for different values of the trade-off parameter $\alpha$ using either oracle diarization labels or estimated diarization labels for GSS \cite{Tawara2025}. When using GSS with oracle diarization labels, $\alpha=0.6$ achieves the lowest macro-average tcpWER, with $\alpha=0.5$ achieving similar performance, while when using GSS with estimated diarization labels, $\alpha=0.5$ achieves the lowest macro-average tcpWER. Therefore, we set $\alpha=0.5$ for all remaining experiments.

Table \ref{tab:oracle_diarization} and Table \ref{tab:estimated_diarization} show the ASR performance of the CHiME-8 system using the baseline and the proposed reference microphone selection methods on the CHiME-8 evaluation data using oracle diarization labels and estimated diarization labels for GSS, respectively. For both oracle and estimated diarization labels, using only the normalized $\ell_p$-norm for reference microphone selection achieves a lower macro-average tcpWER as well as a lower tcpWER in most datasets compared to the baseline method using only the SNR. In addition, using both the normalized $\ell_p$-norm and the SNR further lowers the macro-average tcpWER and achieves a consistent improvement over the baseline method for all datasets using spatially distributed microphones. Note that reference microphone selection does not improve the performance for the NOTSOFAR1 dataset as it does not use spatially distributed microphones. Furthermore, the small improvement seen for the CHiME-6 dataset could be explained by the fact that its linear microphone arrays were distributed in multiple rooms. Therefore, it may be a highly challenging scenario to effectively perform reference microphone selection. Meanwhile, the DiPCo and Mixer 6 datasets had all microphones in the same room. The significant improvement in Mixer 6 could be explained by the fact that it used the highest number of spatially distributed devices.   

\begin{table}[h!]
\centering
\caption{ASR performance of the CHiME-8 system with baseline and proposed reference microphone selection methods in terms of tcpWER (\%) on CHiME-8 evaluation data.}
\label{tab:combined_diarization}
\vspace{-0.2cm}
\begin{subtable}{\columnwidth}
    \centering
    \caption{Oracle diarization}
    \label{tab:oracle_diarization}
    \resizebox{\linewidth}{!}{%
    \begin{tabular}{lccccc}
    \toprule
    \textbf{Method} & \textbf{CH6} & \textbf{DiP} & \textbf{Mi6} & \textbf{NSF} & \textbf{Macro-Average} \\
    \midrule
    SNR & 24.3 & 24.2 & 14.4 & \textbf{13.5} & 19.1 \\
    Normalized $\ell_p$-norm & 24.6 & 23.1 & 13.4 & \textbf{13.5} & 18.7 \\
    Combination ($\alpha=0.5$) & \textbf{24.2} & \textbf{22.9} & \textbf{12.9} & \textbf{13.5} & \textbf{18.4} \\
    \bottomrule
    \end{tabular}}
\end{subtable}

\vspace{0.2cm} 

\begin{subtable}{\columnwidth}
    \centering
    \caption{Estimated diarization}
    \label{tab:estimated_diarization}
    \resizebox{\linewidth}{!}{%
    \begin{tabular}{lccccc}
    \toprule
    \textbf{Method} & \textbf{CH6} & \textbf{DiP} & \textbf{Mi6} & \textbf{NSF} & \textbf{Macro-Average} \\
    \midrule
    SNR & 37.2 & 28.1 & 16.1 & \textbf{20.6} & 25.5 \\
    Normalized $\ell_p$-norm & 37.2 & 26.9 & 13.8 & \textbf{20.6} & 24.6 \\
    Combination ($\alpha=0.5$) & \textbf{37.0} & \textbf{26.7} & \textbf{13.3} & \textbf{20.6} & \textbf{24.4} \\
    \bottomrule
    \end{tabular}}
\end{subtable}

\end{table}

\vspace*{-0.2cm}
\section{Conclusion}
\label{sec:conclusion}
In this paper, we proposed reference microphone selection methods based on the normalized $\ell_p$-norm for GSS-based speech enhancement. We proposed two reference microphone selection methods based on the normalized $\ell_p$-norm, using only the normalized $\ell_p$-norm or combining it with the SNR in order to account for both differences in ELR and SNR across microphones. Experimental evaluating using signal quality metrics showed that our proposed methods select a reference microphone with a higher signal quality compared to the baseline method using only the SNR. Experimental evaluation on the CHiME-8 ASR task showed that the proposed methods achieved a lower macro-average tcpWER compared to the baseline method, with the combined method achieving the lowest tcpWER. 

\bibliographystyle{IEEEbib}
\bibliography{refs}

\end{document}